\documentclass[twocolumn,showpacs,amsmath,amssymb,pre]{revtex4}
\usepackage{graphicx}

\begin{document}

\title{Lyapunov exponent for a gas of soft scatterers}

\author{P.V. Elyutin\footnote {Electronic address:
\texttt{pve@shg.phys.msu.su}}}
\address{Department of Physics, Moscow State University,
Moscow 119992, Russia}

\date{\today}

\begin{abstract}
For a fast particle moving within a two-dimensional array of soft
scatterers - centers of weak and short-range potential - the
dependence of the Lyapunov exponent on the system parameters is
studied.  The use of the linearized equations for variations of
the propagation angles and impact parameters of consequent
collisions reduces the problem to that of calculation of the
Lyapunov exponent of an ensemble of strongly correlated random
matrices with given statistics of matrix elements.  In the
simplest approximation this Lyapunov exponent is proportional to
the interaction strength and inversely proportional to the square
root of the interaction range.  The model satisfactorily describes
the intensity of chaos in a system of two weakly interacting
particles moving in a two-dimensional regular confining potential.

\end{abstract}
\pacs{05.45.-a, 05.45.Pq, 51.10+y} \maketitle

\textbf{1.} The chaoticity of motion of the gas of hard scatterers
- rigid spheres - was established by N.S. Krylov in 1943, twenty
years before the foundation of the paradigm of the modern
nonlinear dynamics \cite{K50}.  He obtained the estimate of the
Lyapunov exponent $\sigma$ that in two-dimensional case has the
form
\begin{equation}
\sigma \sim nav \ln \frac{1}{a\sqrt{n}},
\end{equation}
where $n$ is the concentration of scatterers, $a$ is the sphere
(disk) radius, $v$ is some averaged velocity of the particle and
the condition $a\sqrt{n} \ll 1$ is fulfilled. This result was
refined in the following studies of the model of Lorentz gas of
rigid spheres \cite{vBD95, vBLD98} and confirmed numerically
\cite{vZBD98}.

More realistic case of soft scatterers, in which the interaction
between the particles smoothly depends on the interparticle
distance, is much less clear. The result for this model derived by
Barnett \textit{et al.} \cite{BT+96} later was found to be in
qualitative disagreement with data of numerical experiments
\cite{TDP99}. Recently the case of soft scatterers has been
approached by Kimball \cite{K01}, but the additional superimposed
restrictions has led this author just to rederivation of the
Krylov estimate Eq. (1).

In this letter we study the dependence of the Lyapunov exponent on
parameters of a system of soft scatterers.

\textbf{2.}  We shall use the model of the Lorentz gas - a
particle of mass $m$ moving with velocity $v$ within a
two-dimensional array of fixed potential centers located at the
random points $\textbf{R}_j$ with concentration $n$. The potential
energy of the particle at a point $\textbf{r}$ is
\begin{equation}
V_\Sigma(\textbf{r})=\sum_j V(|\textbf{r}-\textbf{R}_j|),
\end{equation}
where the potential of each center will be taken in the form
$V(r)=V_0f(\alpha r)$ with the function $f(z)$ that rapidly (e.g.
exponentially) decreases with increase of $z$.  In the following
parameters $V_0$ and $\alpha$ will be called the interaction
strength and the inverse interaction range respectively.

First, we confine ourselves to the case of low density, when the
average distance between scattering centers $L \sim n^{-1/2}$ is
much larger than the interaction range: $L \alpha \gg 1$.  Then we
can treat the motion of a particle as a sequence of collisions
with separate centers. The collision with the $i$-th center is
characterized by the propagation angle $\phi_i$ before the
collision and the impact parameter of the collision $\rho_i$.

Next, we assume that the velocity $v$ of the moving particle is
high enough to justify the use of the small-angle approximation,
in which the deflection angle $\theta$ for a collision is given by
the expression \cite{LL88}
\begin{equation}
\theta=-\frac{2\rho}{mv^2}{\int_\rho^\infty\frac{dV}{dr}
\frac{dr}{\sqrt{r^2-\rho^2}}}
=\frac{V_0}{mv^2}\Psi(\alpha\rho).\end{equation}

The linear transformation that relates consequent values of small
variations of the propagation angle and the impact parameter of
the following collision can be written (see \cite{K01}) as
\begin{equation}
\left\{ {\begin{array}{*{20}c}
   \delta\phi_{i+1}  \\
   \delta\rho_{i+1}  \\
\end{array}} \right\}=\hat M_i
\left\{ {\begin{array}{*{20}c}
   \delta\phi_{i}  \\
   \delta\rho_{i}  \\
\end{array}}\right\},
\end{equation}
where $\hat M_i$ is the stability matrix for the $i$-th collision:
\begin{equation}
\hat M_i=\frac{\partial\left(\phi_{i+1},\rho_{i+1}\right)}
{\partial\left(\phi_{i},\rho_{i}\right)}.
\end{equation}
The form of stability matrix $\hat M_i$ depends essentially on
position of consequent scattering centers relatively to the
particle trajectory. This trajectory separates the plane in two
domains; for convenience we shall refer to them as to regions
above and below the trajectory. Then we have four different cases:
1) both scatterers are above the trajectory; 2) the first is
above, the second is below; 3) the first is below, the second is
above and 4) both scatterers are below the trajectory. For these
cases we have respectively
\begin{eqnarray}\nonumber
\hat M_{^i }^{(1)}  = \left| {\begin{array}{*{20}c}
   1 & {\xi _i }  \\
   { - \eta _i } & {1 - \xi _i \eta _i }  \\
\end{array}} \right|,\,\,\,\,\,\,\,
\hat M_{^i }^{(2)}  = \left| {\begin{array}{*{20}c}
   1 & {\xi _i }  \\
   {\eta _i } & { - 1 + \xi _i \eta _i }  \\
\end{array}} \right|,\\ \nonumber
\vspace{3mm}\\
\hat M_{^i }^{(3)}  = \left| {\begin{array}{*{20}c}
   1 & { - \xi _i }  \\
   {\eta _i } & { - 1 - \xi _i \eta _i }  \\
\end{array}} \right|,\,\,\,\,\,\,\,
\hat M_{^i }^{(4)}  = \left| {\begin{array}{*{20}c}
   1 & { - \xi _i }  \\
   { - \eta _i } & {1 + \xi _i \eta _i }  \\
\end{array}} \right|,
\end{eqnarray}
where $\xi_i$ is the derivative of the deflection angle on the
impact parameter for the $i$-th collision,
\begin{equation}
\xi_i=\left. {\frac{\partial \theta}{\partial \rho}}\, \right
|_{\rho=\rho_i}=\left. \frac{V_0 }{mv^2}\, \alpha \,
\frac{d\Psi}{dz}\right|_{\rho=\rho_i},
\end{equation}
and $\eta_i$ is the distance between the $i$-th and the $(i+1)$-th
scattering centers.

The distributions of variables $\xi_i$ and $\eta_i$ can be
characterized by their first moments, the mean values $\tilde
\xi_1=\langle \xi \rangle$, $\tilde \eta_1=\langle \eta \rangle$
and the fluctuations (standard deviations) $\tilde \xi_2=\langle
(\xi-\tilde \xi_1)^2\rangle^{1/2}$, $\tilde \eta_2=\langle
(\eta-\tilde \eta_1)^2\rangle^{1/2}$.  The moments of $\xi$ can be
calculated from the distribution of the impact parameters of
collisions $w(\rho)$. Since this distribution has a characteristic
length about the average distance $L$ between the scatterers, for
$L \gg \alpha^{-1}$ we can replace this distribution by a
constant, $w(\rho)\rightarrow w(0)\sim L^{-1}$, that yields
\begin{equation}
\tilde \xi_1=0, \qquad \tilde
\xi_2=c_1\,\frac{|V_0|}{mv^2}\sqrt{\frac{\alpha}{L}},
\end{equation}
where $c_1$ is a numerical constant of order unity that depends on
the exact form of the interaction potential.  The distribution of
quantities $\eta_i$ does not depend on the interaction parameters.
Its moments have values $\tilde \eta_1=c_2L$ and $\tilde
\eta_2=c_3L$ where $c_3$ and $c_3$ are numerical constants of
order unity that could be found by means of statistical geometry.
From the dimensionality considerations it follows that the
dimensionless Lyapunov exponent "per collision" can depend only on
dimensionless combinations $X=\tilde \xi_2 \tilde \eta_1$ and
$Y=\tilde \xi_2 \tilde \eta_2$.

Although one can assume the values of $\xi_i$ and $\eta_i$ to be
non-correlated, the sequence of types of matrices $\hat M_i$ is
correlated strongly: all matrices have equal probabilities
$p=1/4$, but after any given one only the matrices of two types
can follow with probabilities $p=1/2$. These correlations preclude
the use of the standard analytical methods of calculation the
Lyapunov exponent through the small disorder expansion
\cite{DG84}. We note in passing that for the ensemble of matrices
only of the type $\hat M^{(1)}$ or only of the type $\hat M^{(4)}$
with identical $\eta_i\equiv \tilde \eta_1$ ($\tilde \eta_2 \equiv
0$) an analytical expression for the Lyapunov exponent $\sigma =
0.289 X^{2/3}$ has been found in the context of the theory of
one-dimensional Anderson localization \cite{DG84}.

Function $\sigma(X,Y)$ has been determined by numerical
experiments. The sequence of positions of scattering centers and
values of $\xi_i$ (with gaussian distribution) and $\eta_i$ (with
the Poisson distribution) were created by random number
generators. Then the Lyapunov exponent has been calculated for a
product of $10^5$ matrices.  Comparison of different runs shows
that this volume of computation gives the values of $\sigma$ with
an error $\Delta\sigma \lesssim 5 \cdot 10^{-4}$. The data found
with different parameters (see Fig. 1) are accurately fit by the
simple linear function
\begin{equation}
\sigma = 0.31 X.
\end{equation}

\begin{figure}
[!ht]
\includegraphics[width=0.9\columnwidth]{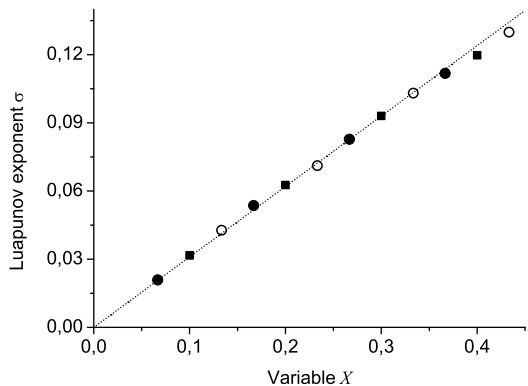}
\caption{\label{fig3} Dependence of the Lyapunov exponent $\sigma$
of the sequence of random matrices $\hat M_i^{(j)}$ given by Eq.
(6) on the variable $X=\tilde \xi_2 \tilde \eta_1$ for different
values of $\tilde \eta_1$: squares for $\tilde \eta_1=1$, filled
circles for $\tilde \eta_1=2$, open circles for $\tilde \eta_1=3$.
The line shows the dependence given by Eq. (9).}
\end{figure}

Now from Eqs. (8) and (9), taking into account the average time
interval between collisions $\Theta \approx L/v$, we obtain for
the Lyapunov exponent of a particle moving within a
two-dimensional array of soft scatterers the expression
\begin{equation}
\sigma = c\,\frac{|V_0|}{mv}\alpha^{1/2} n^{1/4},
\end{equation}
where $c$ is a numerical constant.  This formula presents the main
result of the paper.  The comparison of Eq. (10) with the Krylov
formula Eq. (1) shows the qualitatilevely opposite dependence on
the velocity $v$ and on the interaction range $a \sim \alpha^{-1}$
and quantitatively different dependence on the concentration $n$.

\textbf{3.}  To check the expression Eq. (10) we consider a system
of two identical particles moving in a confining potential
$U(x,y)$. If the motion of a single particle in this potential is
regular, then chaos in two-particle system can originate only from
interaction. If the size $\Lambda$ of the domain of the
configuration space available to particles is much larger then the
interaction range, $\Lambda \alpha \gg 1$, then we can treat the
process of interaction as a series of collisions. In this case we
expect that the Lyapunov exponent will be given by the expression
Eq. (10) appropriately averaged over the distribution of relative
velocities of colliding particles. After this averaging the
dependence $\sigma \propto |V_0|\sqrt{\alpha}$ must still hold.

We checked this dependence by a numerical experiment.  The
confining potential was taken in the form $U(x,y)=K(x^4+y^4)$. The
regular character of motion of a single particle in this potential
is secured by the existence of two integrals of motion, namely
$I_x=m\dot x^2/2+Kx^4$ and $I_y=m\dot y^2/2+Ky^4$.  The
interaction potential was chosen in the form $V(r)=V_0
\cosh^{-1}\alpha r$. We have used the dimensionless units based on
the choice of scales $m=1$, $K=1/4$ and the total energy of the
system $E=40$. The dependences of the Lyapunov exponent on the
interaction strength $V_0$ and on the inverse interaction range
$\alpha$ are shown in Figs. 2 and 3.

\begin{figure}
\includegraphics[width=0.8\columnwidth]{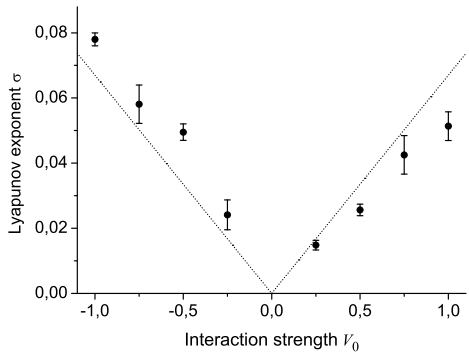}
\caption{\label{fig2} Dependence of the Lyapunov exponent $\sigma$
of a two-particle system in a regular confining potential on the
interaction strength $V_0$ for the total energy $E=40$ and the
inverse interaction range $\alpha=1$. The line shows the fitted
function $\sigma=0.067|V_0|$.}
\end{figure}

\begin{figure}
\includegraphics[width=0.8\columnwidth]{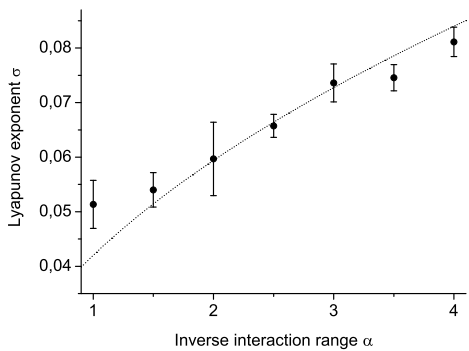}
\caption{\label{fig3} Dependence of the Lyapunov exponent $\sigma$
of a two-particle system in a regular confining potential on the
inverse interaction range $\alpha$ for the total energy $E=40$ and
the interaction strength $V_0=1$. The line shows the fitted
function $\sigma=0.042\sqrt{\alpha}$.}
\end{figure}

The accuracy $\bar \delta$ of agreement between the theory and
numerical calculations for the dependence $\sigma(\alpha)$ is
about 6\%, that is quite satisfactory.  However for the dependence
$\sigma(V_0)$ it is much larger, $\bar \delta=25\%$, and some
explanation is appropriate. It may be observed that Fig. 2
demonstrates a persistent asymmetry: numerical values of $\sigma$
for the attraction ($V_0<0$) happen to be larger (on the average
by a factor about 1.6) than those for the repulsion ($V_0>0$) with
the same $|V_0|$.  In our theory the symmetry in sign of $V_0$
comes as a direct consequence of the small-angle approximation Eq.
(3). From Eq. (10) one can infer importance of collisions with
small relative velocities $v$, for which the small-angle
approximation fails.  It should be noted that arguments for the
asymmetry of the observed type were given by Kimball \cite{K01}
for the case of strong scattering.

In conclusion we note that the model studied in this section can
serve as a classical counterpart to quantum models of two-electron
(or electron-hole) systems confined in quantum dots, that has
attracted much attention lately \cite{LSB00,FST01,VV+01}.

\textbf{4. } The author is grateful to Dr. Shan Jie (Case Western
Reserve University) and to Dr. E.A. Ostrovskaya (Australian
National University) for useful discussions and informational
assistance. Author acknowledges the support by the "Russian
Scientific Schools" program (grant \# NSh - 1909.2003.2).

\end{document}